\documentclass[aps,preprint,floatfix,nofootinbib,showpacs]{revtex4-1}
\pdfoutput=1
\usepackage{graphicx,color,rotating}
\usepackage{hyperref}
\usepackage{epsfig,color}

\newcommand{\lsim}{\raisebox{-0.13cm}{~\shortstack{$<$ \\[-0.07cm] $\sim$}}~}
\newcommand{\gsim}{\raisebox{-0.13cm}{~\shortstack{$>$ \\[-0.07cm] $\sim$}}~}

\def\beq{\begin{equation}}
\def\eeq{\end{equation}}

\def\bea{\begin{eqnarray}}
\def\eea{\end{eqnarray}}
\def\bei{\begin{itemize}}
\def\eei{\end{itemize}}
\def\bmat{\begin{matrix}}
\def\emat{\end{matrix}}
\def\ble{\begin{flushleft}}
\def\ele{\end{flushleft}}
\def\={\,=\,}
\def\+{\,+\,}
\def\-{\,-\,}

\begin{document}

\def\thefootnote{\fnsymbol{footnote}}

{\small
\begin{flushright}
CNU-HEP-15-08
\end{flushright} }

\title{
Higgs precision study of the 750 GeV diphoton resonance and the 125 GeV standard
model Higgs boson with Higgs-singlet mixing
}

\author{
Kingman Cheung$^{1,2,3}$, P. Ko$^4$,
Jae Sik Lee$^{5}$, Jubin Park$^{5}$, and Po-Yan Tseng$^1$}
\affiliation{
$^1$ Department of Physics, National Tsing Hua University,
Hsinchu 300, Taiwan \\
$^2$ Division of Quantum Phases and Devices, School of Physics,
Konkuk University, Seoul 143-701, Republic of Korea \\
$^3$ Physics Division, National Center for Theoretical Sciences, Hsinchu, Taiwan \\
$^4$ School of Physics, KIAS, Seoul 130-722, Republic of Korea \\
$^5$ Department of Physics, Chonnam National University, \\
300 Yongbong-dong, Buk-gu, Gwangju, 500-757, Republic of Korea
}
\date{December 25, 2015}

\begin{abstract}
We interpret the potential observation of the 750 GeV 
di-photon resonance
at the LHC in models, in which an
$SU(2)$ isospin-singlet scalar boson mixes with the Standard Model
(SM) Higgs boson through an angle $\alpha$.  Allowing the singlet
scalar boson to have renormalizable couplings to vector-like leptons
and quarks 
and introducing sizable decay width of the 750 GeV di-photon resonance
into non-SM particles such as dark matters,
we can explain the large production cross section
$\sigma(H_2) \times B(H_2 \to \gamma\gamma)$ as well as the apparent
large total width of the boson
without conflicts from the
results obtained by previous global fits to the SM Higgs boson data.
\end{abstract}

\maketitle

\section{Introduction}

The biggest triumph of the LHC Run I was the discovery of the 
Standard Model (SM) like Higgs boson with mass about 125 GeV \cite{atlas,cms}.
The signal-strength data and the spin-parity of the observed 125 GeV particle  
have all indicated that it is very close to the SM Higgs boson
\cite{Cheung:2013kla,Cheung:2014noa}. After a shutdown for 2 years, the
Run II started with a high expectation. Just with an accumulated luminosity
of about 3 fb$^{-1}$ at $\sqrt{s}=13$ TeV, both ATLAS \cite{atlas2}
and CMS \cite{cms2}
showed a hint of a new particle at about 750 GeV 
decaying into a photon pair. The particle is likely to be a scalar boson
or a spin-2 particle.  We focus on the scalar boson scenario in this paper. 

With a luminosity of 3.2 fb$^{-1}$,  the ATLAS Collaboration found 
a resonance structure at $M_X \approx 750$ GeV with a local significance of 
$\sim 3.64 \sigma$, but corresponding to $1.88\sigma$ when the look-elsewhere-effect is taken into account \cite{atlas2}. 
The CMS Collaboration also reported a similar though smaller 
excess with a luminosity of 2.6 fb$^{-1}$ at $M_X \approx 760 $ GeV with
a local significance of $2.6\sigma$ but a global significance less than
$1.2\sigma$ \cite{cms2}.  Also, in the analysis of ATLAS a total width of
about 45 GeV is preferred \cite{atlas2}.  

These data could be summarized as follows:
\begin{eqnarray}
{\rm ATLAS} &:& M_X = 750 \;{\rm GeV},\;\;\; 
 \sigma_{\rm fit} (pp \to X \to\gamma\gamma) \approx 10 \pm 3\,
{\rm fb};(95\% \;{\rm CL}),\;\; \Gamma_X \approx 45 \,{\rm GeV} \nonumber \\
{\rm CMS} &:& M_X = 760 \,{\rm GeV} ,\;\;\;
 \sigma_{\rm fit} (pp \to X \to\gamma\gamma) \approx 9 \pm 7\,
{\rm fb};(95\% \;{\rm CL}) \nonumber
\end{eqnarray}
The uncertainties shown are $1.96\sigma$ corresponding to 95\% CL.
Note that we estimate the best-fit cross section from the 95\%CL upper
limits given in the experimental paper, by subtracting the ``expected'' 
limit from the ``observed'' limit at $M_X = 750\; (760)$ GeV for ATLAS
(CMS). 

Although this hint for a new resonance is still very preliminary, it has stimulated 
a lot of phenomenological activities, bringing in a number of models for 
interpretation.  The first category is the Higgs-sector extensions, 
including adding singlet Higgs fields \cite{singlet,hidden,dm}, 
two-Higgs-doublet models and the MSSM \cite{2hdm}. 
But in general it fails to explain the large production cross section of 
$pp \to H \to \gamma \gamma$ in the conventional  settings, unless additional 
particles are added, for example,  vector-like fermions 
\cite{2hdm,singlet,hidden,dm}. 
Another  category is the composite models \cite{composite} that naturally 
contain heavy fermions, through which the production and the di-photon decay 
of the scalar boson can be enhanced.  Other possibilities are also entertained, 
such as axion \cite{axion}, sgoldstini \cite{goldstino}, 
radion/dilaton \cite{radion}, and other models \cite{others}.
More general discussion of the di-photon resonance or its
properties can be found in Refs.~\cite{general}.
The generic feature of the suggested interpretations is to enhance
the production cross section of $pp \to H \to \gamma\gamma$, where $H$
is the 750 GeV scalar or pseudo-scalar boson, by additional particles running
in the $H\gamma\gamma$ decay vertex and/or $Hgg$ production vertex.
Another generic feature though not realized in the CMS data is the relatively
broad width of the particle, which motivates the idea that this particle
is window to the dark sector or dark matter \cite{dm,hidden}.

A possible interpretation for this 750 GeV particle can be 
an $SU(2)$ isospin-singlet scalar.
In this interpretation, a general feature is that the singlet $s$ mixes 
with the SM Higgs doublet $H_{\rm SM}$ through an angle $\alpha$
due to the cubic and quartic potential terms such as
$\mu\,s\,H_{\rm SM}^\dagger H_{\rm SM} +\lambda\,s^2\,H_{\rm SM}^\dagger
H_{\rm SM}$.
Further, we note that the singlet may also
have renormalizable couplings to new
vector-like leptons and quarks \cite{Chpoi:2013wga}.
We assume after mixing the lighter boson is the observed SM-like
Higgs boson $H_1$ at 125 GeV while the heavier one $H_2$ is the one hinted at
750 GeV.
Thus, the 750 GeV scalar boson $H_2$ opens the window to another sector 
containing perhaps dark matter (DM) and other exotic particles.

In our previous global fits to the Higgs-portal type models  
with  the SM Higgs mixing with a singlet scalar boson 
with all the Higgs boson data from Run I  \cite{portal}, 
we have constrained the parameter space of a few models with a singlet scalar. 
In the Higgs-portal singlet-scalar models with hidden sector DM, 
there are no new contributions to the $h \gamma\gamma$ and $hgg$ vertices 
beyond the SM contributions, and the mixing angle  $\alpha$ is constrained to 
$\cos\alpha > 0.86$ at 95\% CL. 
However, in those models with vector-like leptons (quarks) the mixing 
angle can be relaxed to $\cos\alpha > 0.83\;(0.7)$ at 95\% CL. 

The implication is that the 750 GeV scalar boson $H_2$ can be produced in 
$gg$ fusion as if it were a 750 GeV SM Higgs boson but with a suppression
factor $\sin^2\alpha$ if there are no vector-like quarks running in the
$H_2 gg$ vertex. Additional contributions arise when there are vector-like
quarks running in the loop. Similarly, the decay of the scalar boson 
$H_2$ behaves like a 750 GeV SM Higgs boson with each partial width 
suppressed by $\sin^2\alpha$ if there are no vector-like leptons or quarks
running in the $H_2 gg$ and $H_2 \gamma\gamma$ vertices. If this is the
case the branching ratio $B(H_2 \to \gamma\gamma) \sim 10^{-6}$, which is too
small to explain the resonance. In this work, we consider vector-like
leptons and vector-like quarks that can enhance the $H_2 \to \gamma\gamma$
decay substantially to give a large production cross section for
$pp\to H_2 \to \gamma\gamma$. 

Vector-like fermions are quite common in a number of extensions of the SM
with various motivations. Although we can introduce vector-like fermions  in an
{\it ad hoc}
and phenomenological way in order to explain the 750 GeV diphoton excess,
their existence can be understood at theoretically deeper levels.
They appear naturally in models with new chiral $U(1)$ gauge symmetries in order
to cancel gauge anomalies \cite{Ko:2016lai,Ko:2016wce,Ko:2016sxg}, in non-Abelian
gauge extensions such as $SU(3)_C \times SU(3)_L \times U(1)_Y$ model (the
so-called
3-3-1 model where gauge anomalies cancel when three generations of fermions are
considered) \cite{Boucenna:2015pav}, or in flavor models for fermion masses and
mixing
\cite{Bonilla:2016sgx}, to name a few explicit models in the context of 750 GeV
diphoton
excess.   In such models, one can in particular  forbid large bare masses of the
vector-like
fermions if they are chiral under this new $U(1)$ gauge symmetries,
and thus motivate their masses fall into the range we need to accommodate the
750 GeV diphoton excess.

In this paper, we interpret the $750$ GeV di-photon resonance by introducing
an $SU(2)$ singlet taking fully account of its mixing with the SM doublet.  
We show that the large production cross section can be explained if
the singlet scalar has
renormalizable couplings to the  vector-like leptons and quarks.
We further show the possibly large total width can be accommodated
if $H_2$ substantially decay into non-SM particles such as dark matters.

The organization is as follows. In the next section, we describe briefly
the framework of the SM Higgs mixing with a singlet scalar that couples to 
new vector-like fermions.   In Sec. III, we present the numerical results for 
the 750 GeV resonance including the constraints from the properties of 
the 125 GeV SM Higgs-like scalar boson.  Then we conclude in Sec. IV. 

\section{Higgs-singlet mixing framework}
If there are extra vector-like fermions with renormalizable couplings to a singlet 
scalar $s$ 
\footnote{This singlet scalar $s$ could be a remnant of new gauge symmetry
breaking. In that case, $s$ may carry a new quantum number 
different from the SM gauge charges~\cite{mosc}.},  
these models generically 
contain two interaction eigenstates states of $h$ denoting the remnant of 
the SM Higgs doublet and $s$ the singlet.
The two mass eigenstates $H_{1,2}$ are related to the states $h$ and $s$
through an $SO(2)$ rotation as follows:
\begin{equation}
H_1  =  h\,\cos\alpha - s\,\sin\alpha\,; \ \ \ \
H_2  =  h\,\sin\alpha + s\,\cos\alpha\,
\end{equation}
with $\cos\alpha$ and $\sin\alpha$ describing
the mixing between the interaction eigenstates $h$ and $s$.
In the limit of $\sin\alpha\to 0$, $H_1\,(H_2)$ becomes the pure doublet
(singlet) state.
In this work,  we are taking $H_1$ for the 125 GeV boson
discovered at the 8-TeV LHC run
and $H_2$ for the 750 GeV state hinted at the early 13-TeV LHC run.
We are taking $\cos\alpha > 0$ without loss of
generality.
For the detailed description of this class of models and also   Higgs-portal 
models, we refer to  Refs.~\cite{Chpoi:2013wga,portal}.

In this class of models, the singlet field $s$ does not directly couple to the SM 
particles, but only through the mixing with the SM Higgs field at renormalizable 
level. And the Yukawa interactions of $h$ and $s$ are described by
\begin{equation}
-{\cal L}_Y=h\sum_{f=t,b,\tau}\frac{m_f}{v}\bar{f}{f}
+s\sum_{F=Q,L}\,g^S_{s\bar{F}F}\bar{F}F\,,
\end{equation}
with $f$ denoting the 3rd-generation SM fermions and $F$ the extra
vector-like fermions (VLFs):
vector-like quarks (VLQs) and vector-like leptons (VLLs). 
Then the couplings of the
two mass eigenstates  $H_{1,2}$ to the SM and extra fermions 
are given by
\begin{eqnarray}
-{\cal L}_Y&=&
H_1\,\left[\cos\alpha\,\sum_{f=t,b,\tau}\frac{m_f}{v}\bar{f}{f}
-\sin\alpha\,\sum_{F=Q,L}\,g^S_{s\bar{F}F}\bar{F}F\right]
\nonumber \\
&+&
H_2\,\left[\sin\alpha\,\sum_{f=t,b,\tau}\frac{m_f}{v}\bar{f}{f}
+\cos\alpha\,\sum_{F=Q,L}\,g^S_{s\bar{F}F}\bar{F}F\right]\,.
\end{eqnarray}

The couplings of $H_{1,2}$ to two gluons, following the conventions and normalizations of Ref.~\cite{Lee:2003nta}, are given by
\begin{eqnarray}
S_{H_1}^g &=& \cos\alpha\,S^{g\,({\rm SM})}_{H_1}-\sin\alpha\,S^{g\,(Q)}_{H_1}
\nonumber \\
&\equiv &
\cos\alpha\,\sum_{f=t,b}\, F_{sf}(\tau_{1f})-
\sin\alpha\,\sum_Q\,g^S_{s\bar{Q}Q}\,\frac{v}{m_Q}\, F_{sf}(\tau_{1Q})\,,
\nonumber \\
S_{H_2}^g &=& \sin\alpha\,S^{g\,({\rm SM})}_{H_2}
+\cos\alpha\,S^{g\,(Q)}_{H_2} \nonumber \\
&\equiv &
\sin\alpha\,\sum_{f=t,b}\, F_{sf}(\tau_{2f})+
\cos\alpha\,\sum_Q\,g^S_{s\bar{Q}Q}\,\frac{v}{m_Q}\, F_{sf}(\tau_{2Q})\,,
\end{eqnarray}
where $\tau_{ix}=M_{H_i}^2/4m_x^2$.
We note that
$S^{g\,({\rm SM})}_{H_1}\simeq 0.651+0.050\,i$ for $M_{H_1}=125.5$ GeV
and
$S^{g\,({\rm SM})}_{H_2}\simeq 0.291+0.744\,i$ for $M_{H_2}=750$ GeV.
In the limit $\tau\to 0$, $F_{sf}(0)=2/3$.
The mass of extra fermion $F$ may be fixed by the relation
$m_F=v_s\,g^S_{s\bar{F}F}+m^0_F$ where $v_s$ denotes the VEV of the singlet $s$
while  $m^0_F$ is generated from a different origin other than $v_s$ as in
$-{\cal L}_{\rm mass}\supset m^0_F\bar{F}F$.
We note that, when $m^0_Q=0$, each contribution from a VLQ is not 
suppressed by $1/m_Q$ but by the common factor $1/v_s$.

Similarly, the couplings of $H_{1,2}$ to two photons are given by
\begin{eqnarray}
S_{H_1}^\gamma &=& \cos\alpha\,S^{\gamma\,({\rm SM})}_{H_1}-
\sin\alpha\,S^{\gamma\,(F)}_{H_1}
\nonumber \\
&\equiv &
\cos\alpha\left[2\sum_{f=t,b,\tau}\, N_CQ_f^2 F_{sf}(\tau_{1f})
-F_1(\tau_{1W})\right]-
\sin\alpha\left[2\sum_F\,N_CQ_F^2 g^S_{s\bar{F}F}\,\frac{v}{m_F}\,
F_{sf}(\tau_{1F})\right]\,,
\nonumber \\
S_{H_2}^\gamma &=& \sin\alpha\,S^{\gamma\,({\rm SM})}_{H_2}
+\cos\alpha\,S^{\gamma\,(F)}_{H_2} \nonumber \\
&\equiv &
\sin\alpha\left[2\sum_{f=t,b,\tau}\,N_CQ_f^2 F_{sf}(\tau_{2f})
-F_1(\tau_{2W})\right]+
\cos\alpha\left[2\sum_F\,N_CQ_F^2 g^S_{s\bar{F}F}\,\frac{v}{m_F}\,
F_{sf}(\tau_{2F})\right]\,,\nonumber\\
\end{eqnarray}
where $N_C=3$ and $1$ for quarks and leptons, respectively, and
$Q_{f,F}$ denote the electric charges of fermions in the unit of $e$.
In the limit $\tau\to 0$, $F_1(0)=7$.
We note that
$S^{\gamma\,({\rm SM})}_{H_1}\simeq -6.55+0.039\,i$ for $M_{H_1}=125.5$ GeV
and
$S^{\gamma\,({\rm SM})}_{H_2}\simeq -0.94-0.043\,i$ for $M_{H_2}=750$ GeV.

The production cross section of $H_2$ via the gluon-fusion process 
is given by
\begin{equation}
\label{eq:sigh2}
\sigma(gg\to H_2) = 
\frac{|S^g_{H_2}|^2}{|S^{g\,({\rm SM})}_{H_2}|^2}\,\sigma_{\rm SM}(gg\to H_2)
\end{equation}
with
$\sigma_{\rm SM}(gg\to H_2) \approx 800$ fb denoting
the corresponding SM cross section for $M_{H_2}=750$ GeV
at $\sqrt{s}=13$ TeV~\cite{handbook}.
Note that the relation in Eq.~(\ref{eq:sigh2}) only holds at 
leading order.

The total decay width of $H_2$ can be cast into the form
\begin{equation}
\Gamma(H_2)=\sin^2\alpha\,\Gamma_{\rm SM}(H_2)
+\Delta\Gamma_{\rm vis}^{H_2}
+\Delta\Gamma_{\rm inv}^{H_2}\,,
\end{equation}
where
$\Gamma_{\rm SM}(H_2) \simeq 250$ GeV for the SM-like $H_2$
with $M_{H_2}=750$ GeV
\footnote{For $M_{H_2}=750$ GeV,
$\Gamma_{\rm SM}(H_2\to WW)\simeq 145$ GeV,
$\Gamma_{\rm SM}(H_2\to ZZ)\simeq 71.9$ GeV, and
$\Gamma_{\rm SM}(H_2\to t\bar{t})\simeq 30.6$ GeV.
\cite{Dittmaier:2011ti}.}.  
And $\Delta\Gamma_{\rm vis}^{H_2}$ and $\Delta\Gamma_{\rm inv}^{H_2}$ 
denote  additional partial decay widths of $H_2$ into visible and invisible 
particles, respectively.  The quantity $\Delta\Gamma_{\rm vis}^{H_2}$ includes 
the decays into $H_1 H_1$ by definition and, if it is allowed kinematically, 
into extra vector-like fermions as well as those into $\gamma\gamma$, $gg$ through 
the one-loop processes  induced by  the extra VLQs and/or VLLs.
The quantity $\Delta\Gamma_{\rm inv}^{H_2}$ may include the $H_2$ decay into
invisible particles such as dark matters,
or $H_2$ decays into  a pair of Nambu-Goldstone bosons such 
as Majorons which appear in models for neutrino mass generations 
(see Refs.~\cite{Bonilla:2015uwa,Bonilla:2015jdf} for example), or dark 
radiation  (or fractional cosmic neutrinos) which appear when global dark $U(1)$
symmetry is spontaneously broken \cite{Weinberg:2013kea}.

The partial decay width of $H_2$ into two photons is given by
\begin{equation}
\Delta\Gamma^{H_2\to\gamma\gamma}_{\rm vis}
=\frac{M_{H_2}^3\alpha^2}{256\pi^3v^2}
\left[\left|S_{H_2}^\gamma \right|^2
-\sin^2\alpha \left|S_{H_2}^{\gamma\,({\rm SM})} \right|^2\right]
\end{equation}
and that into two gluons is
\begin{equation}
\Delta\Gamma^{H_2\to gg}_{\rm vis}
=\left[1+\frac{\alpha_s}{\pi}\left(\frac{95}{4}-7\right)\right]\,
\frac{M_{H_2}^3\alpha_s^2}{32\pi^3v^2}
\left[\left|S_{H_2}^g \right|^2
-\sin^2\alpha \left|S_{H_2}^{g\,({\rm SM})} \right|^2\right]
\end{equation}
with $\alpha_s=\alpha_s(M_{H_2})$.
%


\section{Numerical Results}
In our numerical analysis, we shall restrict ourselves to the case 
$2m_F>M_{H_2}$ so that  $H_2\to F\bar{F}$ decays are kinematically forbidden 
and  $S^{g(Q),\gamma(F)}_{H_1,H_2}$ are all real.
In this case, one may carry out a model-independent study on
the 750 GeV di-photon resonance with the following varying parameters:
\begin{equation}
\sin\alpha\,, \ \ \
S_{H_2}^{g(Q)}\,, \ \ \
S_{H_2}^{\gamma(F)}\,, \ \ \
\Gamma_{H_2}^{\rm non-SM}\,, \ \ \ 
\eta^{g(Q)}\,, \ \ \
\eta^{\gamma(F)}\,,
\end{equation}
where
\begin{equation}
\Gamma_{H_2}^{\rm non-SM}\equiv
\Gamma(H_2 \to H_1H_1) + 
\Delta\Gamma^{H_2}_{\rm inv}\,.
\end{equation}
%
Here the parameters $\eta^{g(Q)}$ and $\eta^{\gamma(F)}$ are defined as in
\begin{equation}
S_{H_1}^{g(Q)} \equiv \eta^{g(Q)} S_{H_2}^{g(Q)}\,, \ \ \
S_{H_1}^{\gamma(F)} \equiv \eta^{\gamma(F)} S_{H_2}^{\gamma(F)}\,.
\end{equation}
We note that $\eta^{g(Q)}$  and $\eta^{\gamma(F)}$ take values between
$2/3$ and $1$ for the following reasons: 
\begin{eqnarray}
&&
S_{H_1}^{g(Q)}   
=\sum_Q g^S_{s\bar{Q}Q}\frac{v}{m_Q}F_{sf}(\tau_{1Q})
\simeq\frac{2}{3}\sum_Q g^S_{s\bar{Q}Q}\frac{v}{m_Q}\,,
\nonumber \\[2mm]
&&
\frac{2}{3} \sum_Q g^S_{s\bar{Q}Q}\frac{v}{m_Q}
\leq S_{H_2}^{g(Q)}
=\sum_Q g^S_{s\bar{Q}Q}\frac{v}{m_Q}F_{sf}(\tau_{2Q})
\leq \sum_Q g^S_{s\bar{Q}Q}\frac{v}{m_Q}\,,
\end{eqnarray}
if we have $g^S_{s\bar{Q}Q}>0$ for all $Q$'s 
\footnote{In this study, we take the more conventional choice of 
$g^S_{s\bar F F} > 0$  for the Yukawa-type coupling between $s$ and VLFs. 
In general, it may be  possible to have negative $g^S_{s\bar F F}$ for 
some VLFs in specific models
and the parameters $\eta^{g(Q),\gamma(F)}$  can take any values in principle.
However, we shall fully investigate such a case in
a later work~\cite{DD}.}.  

Since $|S^{g(Q),\gamma(F)}_{H_1}|$ is larger than
$\frac{2}{3}\,|S^{g(Q),\gamma(F)}_{H_2}|$, 
the parameters $S^{g(Q),\gamma(F)}_{H_2}$
can not be arbitrarily large without affecting the LHC data on 125 GeV
Higgs boson when $\sin\alpha\neq 0$. For example, the quantities
\begin{equation}
C_{H_1}^{g\,,\gamma} = |S_{H_1}^{g\,,\gamma}|/|S_{H_1}^{g\,,\gamma(\rm SM)}|
\end{equation}
can not significantly deviate from $1$~\cite{Cheung:2013kla}.
If $\sin\alpha\, |S_{H_1}^{g(Q)}|$ and $\sin\alpha\, |S_{H_1}^{\gamma(F)}|$
are required to be within the $\pm 10$\% range
of the corresponding SM values, one might have
\begin{equation}
|S_{H_2}^{g(Q)}|\lsim \frac{0.1}{|\sin\alpha|}\,, \ \ \
|S_{H_2}^{\gamma(F)}|\lsim \frac{1}{|\sin\alpha|}\,,
\end{equation}
when $\eta^{g(Q)}=\eta^{\gamma(F)}=2/3$.
Therefore, we again restricted ourselves to
the case of $|\sin\alpha|\lsim 0.1$ in order to have 
$|S_{H_2}^{g(Q)}| \gsim {\cal O}(1)$  and  
$|S_{H_2}^{\gamma(F)}| \gsim {\cal O}(10)$.

When $\sin\alpha \sim 0$, we have numerically
\begin{eqnarray}
\label{eq:numeric}
\sigma(gg\to H_2) &\sim & 1250\, |S^{g(Q)}_{H_2}|^2  \ {\rm fb}
\,, \nonumber \\[2mm]
\Gamma(H_2 \to \gamma\gamma) &\sim & 4.67\times 10^{-5}\,
|S^{\gamma(F)}_{H_2}|^2  \ {\rm GeV}
\,, \nonumber \\[2mm]
\Gamma(H_2 \to gg) &\sim & 8.88\times 10^{-2}\,|S^{g(Q)}_{H_2}|^2  \ {\rm GeV}
\,, \nonumber \\[2mm]
\sigma(gg\to H_2)\times B(H_2 \to \gamma\gamma) &\sim &
11.8\,\frac{\left(|S^{g(Q)}_{H_2}S^{\gamma(F)}_{H_2}|/90\right)^2}
{\left(\Gamma_{H_2}/40\,{\rm GeV}\right)}\,{\rm fb}
\end{eqnarray}
where $\Gamma_{H_2}\sim \Gamma(H_2 \to \gamma\gamma)
+\Gamma(H_2 \to gg) + \Gamma(H_2 \to H_1H_1) + 
\Delta\Gamma^{H_2}_{\rm inv}$.

First of all, to have $\Gamma(H_2 \to \gamma\gamma) \sim 40$ GeV, one
needs $|S^{\gamma(F)}_{H_2}|^2 \sim 10^6$ which requires 
unlikely large value of $Q_F \gsim 10$ with 
$g^S_{s\bar{F}F}\sim 1$ and $m_F=400$-$500$ GeV. If $Q_F \sim {\cal O}(1)$,
$\Gamma(H_2 \to \gamma\gamma)$ is significantly smaller than $1$ GeV
since $|S^{\gamma(F)}_{H_2}|^2 \propto Q_F^4$.
On the other hand,
to have $\Gamma(H_2 \to gg)\sim 40$ GeV, one needs 
$|S^{g(Q)}_{H_2}|^2 \sim 4\times 10^2$ which 
results in $\sigma(gg\to H_2)\sim 5\times 10^5$ fb
leading to enormous number of di-jet events with $B(H_2\to gg) \sim 1$.
Therefore, one may need to have 
\begin{equation}
\Gamma_{H_2}\sim
\Gamma_{H_2}^{\rm non-SM}
\sim 40~{\rm GeV}\,.
\end{equation}

Secondly, we note that $|S^{g(Q)}_{H_2}S^{\gamma(F)}_{H_2}|\sim 90$ to accommodate
$\sigma(gg\to H_2)\times B(H_2 \to \gamma\gamma) \sim 10$ fb.
Our representative choice of $S^{g(Q)}_{H_2}=3$ can be easily realized
if there are about 6 VLQs with
$m_Q \sim 400$-$500$ GeV and $g^S_{s\bar{Q}Q} \sim 1$.
Usually $|S^{\gamma(F)}_{H_2}|$ is larger than $|S^{g(Q)}_{H_2}|$ 
enhanced by the $2N_CQ_F^2$ factor together with additional contributions
from VLLs. Therefore $S^{\gamma(F)}_{H_2}=10\times S^{g(Q)}_{H_2}=30$
could be a reasonable choice.

\begin{figure}[t!]
\centering
\includegraphics[height=5.0in,angle=0]{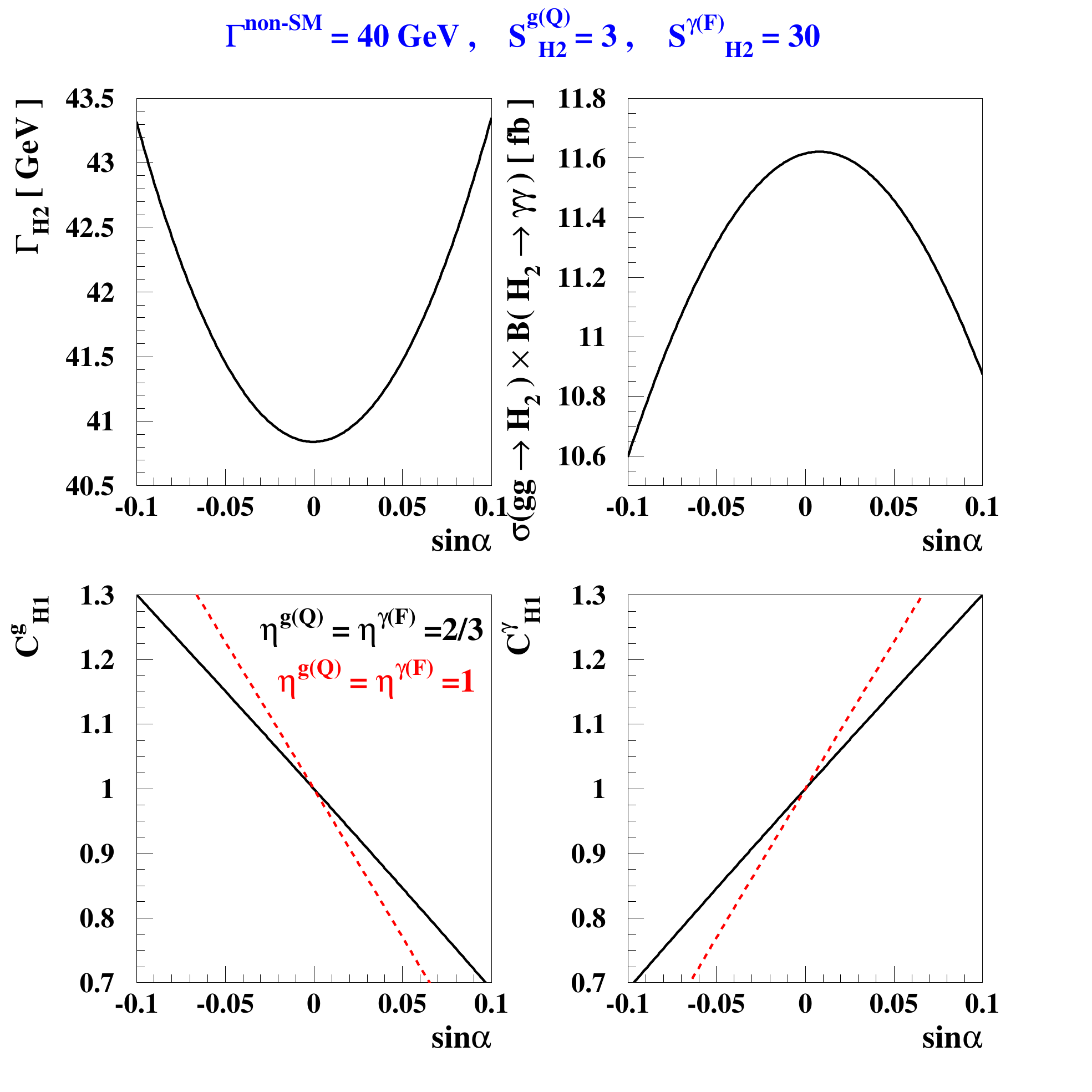}
\caption{\small \label{fig:xxx} 
The decay width $\Gamma_{H_2}$ (upper left),
$\sigma(gg\to H_2)\times B(H_2 \to \gamma\gamma)$ (upper right),
and the ratios $C^{g,\gamma}_{H_1}$ (lower) as functions of $\sin\alpha$.
We have taken $S^{g(Q)}=3$, $S^{\gamma(F)}=10\times S^{g(Q)}=30$,
and $\Gamma_{H_2}^{\rm non-SM} =40$ GeV.
In the lower frames, the solid (dashed) lines are 
for $\eta^{g(Q)}=\eta^{\gamma(F)}=2/3\,(1)$.  }
\end{figure}
Bearing all theses observations, in Fig.~\ref{fig:xxx},
we show
the decay width $\Gamma_{H_2}$ (upper left), the cross section
$\sigma(gg\to H_2)\times B(H_2 \to \gamma\gamma)$ (upper right),
and the ratios $C^{g,\gamma}_{H_1}$ (lower) as functions of $\sin\alpha$
taking
$S^{g(Q)}=3$, $S^{\gamma(F)}=10\times S^{g(Q)}=30$, and $\Gamma_{H_2}^{\rm non-SM}
=40$ GeV.  In the lower frames, the solid (dashed) lines are
for $\eta^{g(Q)}=\eta^{\gamma(F)}=2/3\,(1)$.
We observe that the suggested scenario comfortably explains the properties of
the 750 GeV di-photon resonance without any conflict with the precision data on 125 GeV
Higgs.
A full model-independent precision analysis of the 125-GeV Higgs and
750-GeV resonance data
is to be addressed in a future publication~\cite{DD}.

\begin{figure}[t!]
\centering
\includegraphics[height=5.0in,angle=0]{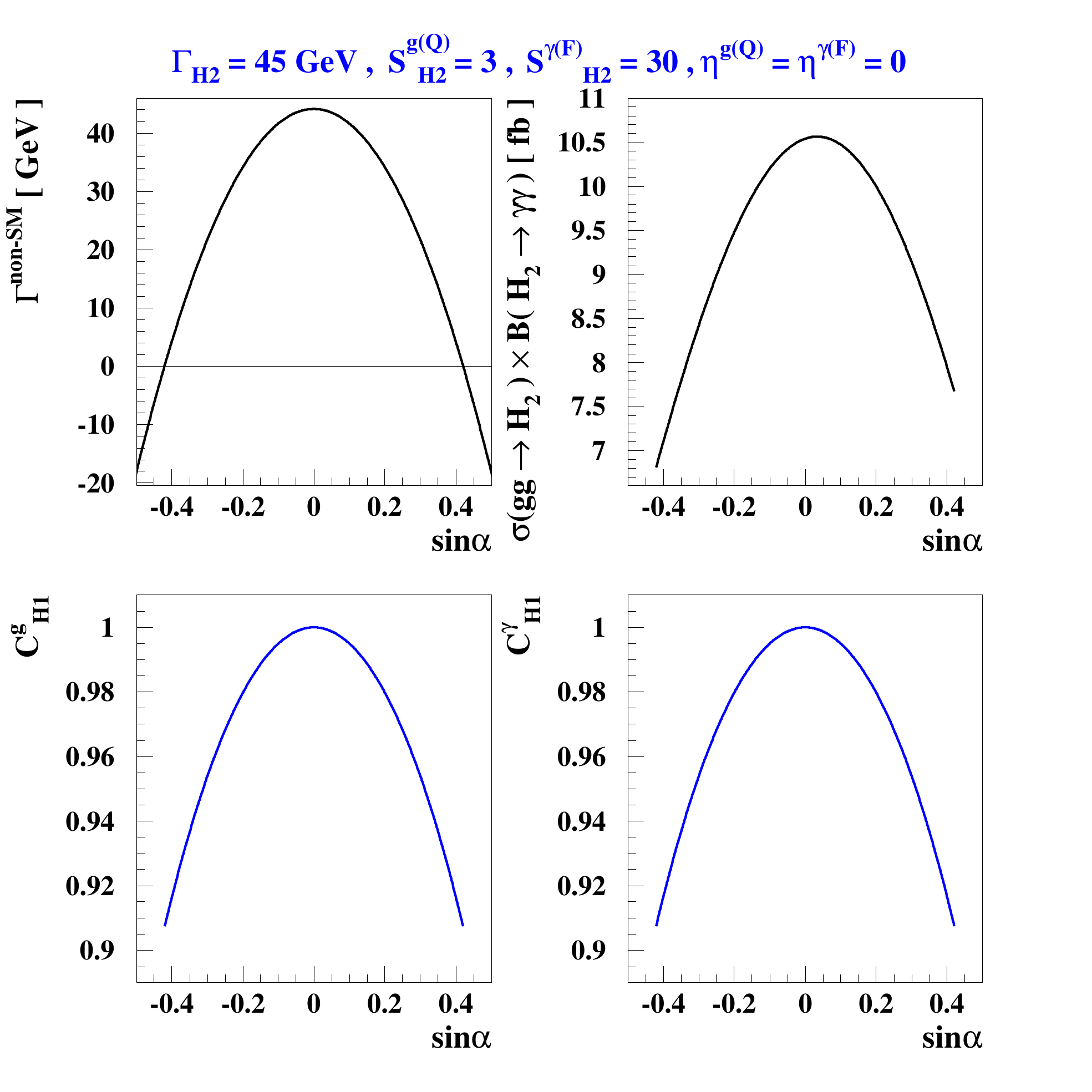}
\caption{\small \label{fig:yyy} 
The non-SM decay width $\Gamma_{H_2}^{\rm non-SM}$ (upper left),
$\sigma(gg\to H_2)\times B(H_2 \to \gamma\gamma)$ (upper right),
and the ratios $C^{g,\gamma}_{H_1}$ (lower) as functions of $\sin\alpha$.
We have taken $S^{g(Q)}_{H_2}=3$, $S^{\gamma(F)}_{H_2}=30$,
and $\Gamma_{H_2} =45$ GeV.
In the upper-right and lower frames, the physical condition
$\Gamma^{\rm non-SM}_{H_2}>0$ is imposed.
}
\end{figure}
Though we have concentrated on the case of $|\sin\alpha|<0.1$, we find
our solution with $S^{g(Q)}=3$ and $S^{\gamma(F)}=30$ 
remains to be valid up to $|\sin\alpha| \sim 0.4$
which is still allowed according to
our global fits to the Higgs-portal type models~\cite{portal}, 
see Fig~\ref{fig:yyy}.
We fix $\Gamma_{H_2}=45$ GeV and tune $\Gamma^{\rm non-SM}_{H_2}$ 
to accommodate it. And a general possibility of having
$\eta^{g(Q))}=\eta^{\gamma(F)}=0$ is considered to satisfy the 
results of the global fits to the 125 GeV Higgs boson data. 
In this case, we note that 
$\sigma(gg\to H_2)\times B(H_2 \to \gamma\gamma) \propto \cos^4\alpha$
and
$C^{g,\gamma}_{H_1}=\cos\alpha$.

In the following, we would like to comment on $H_2$ decays into
$WW,ZZ,t\bar{t}$, and $gg$. 
First, let us consider the case where $H_2$ is produced through the 
SM-singlet VLQs which only have couplings to $g$ and $\gamma$.
In this limit of no interactions between VLQs with the $W/Z$ boson,
$H_2$ decays into $WW$, $ZZ$, and $t\bar{t}$ through its SM Higgs component
at the tree level while the decay into $gg$ proceeds through
the VLQ loops. In this case,
the cross section times branching ratios are
\begin{eqnarray}
\sigma(gg\to H_2)\times B(H_2\to WW) &\simeq & 400\,{\rm fb}\,
\left(\frac{S^{g(Q)}_{H_2}}{3}\right)^2
\left(\frac{\sin\alpha}{0.1}\right)^2
\left(\frac{40\,{\rm GeV}}{\Gamma_{H_2}}\right)
\left(\frac{\sigma_{\rm SM}(gg\to H_2)}{800\,{\rm fb}}\right)
\,, \nonumber\\[2mm]
\sigma(gg\to H_2)\times B(H_2\to ZZ) &\simeq & 200\,{\rm fb}\,
\left(\frac{S^{g(Q)}_{H_2}}{3}\right)^2
\left(\frac{\sin\alpha}{0.1}\right)^2
\left(\frac{40\,{\rm GeV}}{\Gamma_{H_2}}\right)
\left(\frac{\sigma_{\rm SM}(gg\to H_2)}{800\,{\rm fb}}\right)
\,, \nonumber\\[2mm]
\sigma(gg\to H_2)\times B(H_2\to t\bar{t}) &\simeq & 90\,{\rm fb}\,
\left(\frac{S^{g(Q)}_{H_2}}{3}\right)^2
\left(\frac{\sin\alpha}{0.1}\right)^2
\left(\frac{40\,{\rm GeV}}{\Gamma_{H_2}}\right)
\left(\frac{\sigma_{\rm SM}(gg\to H_2)}{800\,{\rm fb}}\right)
\,, \nonumber\\[2mm]
\sigma(gg\to H_2)\times B(H_2\to gg) &\simeq & 200\,{\rm fb}\,
\left(\frac{S^{g(Q)}_{H_2}}{3}\right)^4
\left(\frac{40\,{\rm GeV}}{\Gamma_{H_2}}\right) 
\left(\frac{\sigma_{\rm SM}(gg\to H_2)}{800\,{\rm fb}}\right)
\,.
\end{eqnarray}
Given that the current upper limits on production of a resonance into a 
$ZZ$, $WZ$ , or $WW$
pair is about $150-200$ fb for $M_X =750$ GeV at $\sqrt{s}=13$
TeV \cite{atlas3}, our scenario is more or less safe if 
$|\sin\alpha| \lsim 0.1$.
At $\sqrt{s}=13$ TeV, the search for di-jet resonances did not cover the di-jet
mass range below 1 TeV, and we did not find any search for $t\bar t$
resonances.

On the other hand, at $\sqrt{s}=8$ TeV, the gluon-fusion
production cross section for a SM Higgs boson of 750 GeV is about 150 fb
\cite{higgs-x}.
A combined search for $WW,WZ,ZZ$ resonances has placed at upper limit of
$\sigma(pp\to G^*)\times B(G^* \to VV)$ at slightly less than 100 fb
for $M_{G^*} \approx 750$ GeV \cite{atlas4}.
Therefore, the parameter regions of 
$|\sin\alpha| \alt 0.1$ are perfectly safe
with this 8 TeV search.
Another search for $t\bar t$ resonances put an upper limit of
$\sigma(pp \to X)\times B(X \to t\bar t)$ at about $0.5-1$ pb for
a few models \cite{atlas5}, which is again very safe for our scenario.
Yet, another search for di-jet resonances \cite{atlas6} only covered the
mass range from 0.85 TeV and up. At 0.85 TeV, the production rate limit
is $1-2$ pb, which hardly affects our scenario.

In general there can exist interactions between VLFs and $W/Z$
bosons. To be specific, we consider the case in which VLQs share 
the SM $SU(2)$ and $U(1)_Y$ interactions. 
Then,
in the limit of very small $\sin\alpha$, 
the decay of $H_2$ into $WW$
as well as those into $ZZ$, $Z\gamma$ and $\gamma\gamma$
are dominated by the loops of VLQs.  
These loop-induced decay modes,
especially the $WW$ mode, are more model dependent than 
those into two gluons and two photons and
we consider two representative scenarios for the interactions
between VLQs with $W/Z$ bosons.

In the scenario where VLQs are $SU(2)$ singlets with only
hypercharge interactions, they do not couple to the $W$ boson. While
their interactions with the photon and the $Z$ boson are described by
\begin{equation}
{\cal L}_{VLQ} = - e Q_{VLQ}\, \bar Q \gamma^\mu Q\; A_\mu
       - \frac{e}{s_W c_W}\, (-Q_{VLQ} s^2_W )\,
      \bar Q \gamma^\mu Q\; Z_\mu  \;,
\end{equation}
where we are taking $e>0$ with
$s_W\equiv\sin\theta_W$, $c_W\equiv\cos\theta_W$, and $t_W=s_W/c_W$.
We find that
the effective vertices involving $H_2 \gamma\gamma$, $H_2 Z \gamma $,
and $H_2 Z Z$ can be written as, up to an overall constant,
\begin{equation}
{\cal L} \propto H_2 \left( F_{\mu\nu} F^{\mu\nu} +
   t_W F_{\mu\nu} Z^{\mu\nu} + t^2_W Z_{\mu\nu} Z^{\mu\nu}
   \right ) \;,
\end{equation}
and the ratio $\Gamma(H_2 \to ZZ):\Gamma(H_2 \to Z \gamma):
\Gamma(H_2 \to \gamma\gamma)$ is approximately given by
\begin{equation}
\Gamma(H_2 \to ZZ):\Gamma(H_2 \to Z \gamma):
\Gamma(H_2 \to \gamma\gamma) \,\approx\,
t^4_W : 2 t^2_W : 1 \;,
\end{equation}
ignoring the $Z$-boson mass in the final state. Taking
$t_W \approx 0.55$, the ratio is $0.09: 0.6 :1$.
For $\sigma(pp\to H_2 \to \gamma\gamma)\sim 10$ fb, we have
$\sigma (pp \to H_2 \to ZZ)  \approx  0.9$ fb and
$\sigma (pp \to H_2 \to Z\gamma) \approx  6$ fb which correspond to
$1.4$ $ZZ$ events and $130$ $Z\gamma$ events using $Z\to \ell^+\ell^-$
with an accumulated luminosity of 300 fb$^{-1}$ in the future LHC.

In another scenario, we place one pair of VLQs $U$ and $D$ in an $SU(2)$ doublet
as $(U,D)^T=(U,D)_L^T+(U,D)_R^T$ which carries hypercharge $Y$. Then
the electric charges are given by $Q_U=T_{3U}+Y$ and $Q_D=T_{3D}+Y$
and we have $Q_U - Q_D = 1$ independently of the hypercharge $Y$.
Note we are taking $T_{3U}= -T_{3D} =1/2$.  In this case the interactions
of the VLQs with gauge bosons are given by
\begin{eqnarray}
 {\cal L}_{VLQ} &=& -e \left( Q_U \bar U \gamma^\mu U + Q_D \bar D \gamma^\mu D
   \right )\; A_\mu  \nonumber \\
 &&- \frac{e}{s_W c_W} \,
\left[
   \bar U \gamma^\mu U ( T_{3U} - Q_U s^2_W )
+  \bar D \gamma^\mu D ( T_{3D} - Q_D s^2_W ) \right ]\, Z_\mu
   \nonumber \\
 &&- \frac{e}{\sqrt{2} s_W } \, \left( \bar U \gamma^\mu D W^+_\mu
                                + \bar D \gamma^\mu U W^-_\mu  \right )\,.
\end{eqnarray}
We note the couplings to the $Z$ boson are
purely vector-like and
proportional to the factors $T_{3U,3D} - Q_{U,D} s^2_W$ 
which are different from the SM case where
only the left-handed quarks are participating in the $SU(2)$ interaction. 
It is possible to make a precise prediction in a simpler case in which,
for example, $Y=0$
\footnote{
We find a complete agreement between
our results and those presented in Ref.~\cite{aggkmz}. A more detailed
study considering various scenarios will be presented in Ref.~\cite{DD}.}:
\begin{eqnarray}
&&
\Gamma(H_2 \to WW) : \Gamma(H_2 \to ZZ):\Gamma(H_2 \to \gamma Z):
\Gamma(H_2 \to \gamma\gamma) \,\approx\,
\frac{1}{2 s_W^4 (Q_U^2 + Q_D^2)^2 } :
  \frac{1}{t^4_W}
  : \frac{2}{t^2_W} : 1 \;,
\nonumber \\ &&
\end{eqnarray}
ignoring the $W$- and $Z$-boson masses. Taking $s_W^2 \approx 0.23$
and $Q_{U,D}^2=1/4$, we find the ratio is $ 38 : 11:  6.6 :1$.
For $\sigma(pp\to H_2 \to \gamma\gamma)\sim 10$ fb, we have
$\sigma (pp \to H_2 \to WW)  \approx  380$ fb,
$\sigma (pp \to H_2 \to ZZ)  \approx  110$ fb, and
$\sigma (pp \to H_2 \to Z\gamma) \approx  66$ fb which correspond to
$5400$ $WW$ events, $180$ $ZZ$ events, and $1400$ $Z\gamma$ events 
using $Z\to \ell^+\ell^-$ and $W\to \ell \nu$
with an accumulated luminosity of 300 fb$^{-1}$ in the future LHC.
This scenario is much more promising to probe compared to the previous one

Before concluding, we would like to make a comment on the LHC constraints 
on VLQs.
The VLQs have been actively searched for at the LHC. For example,
the ATLAS and CMS collaborations carried out searches recently
at $\sqrt{s}=13$ TeV \cite{vlq-13,vlq-cms} and there was
another one at 8 TeV \cite{vlq-8}. The lower limits on VLQ mass range
from about 750 GeV to about 1.7 TeV, depending on decay channels.
Such channels include ${\rm VLQ}\to b W, Z t, H t$.
Note that all the particles in the final states are visible and energetic 
because the mass differences between the
VLQ and decay products are assumed to be large enough.
Furthermore, the branching ratio into a chosen decay channel is assumed 100\%.
However, if the VLQ decays into invisible particles, e.g., dark matter,
and other SM particles, and also if the mass difference between the VLQ
and dark matter is small, then the energy available for the visible
particles would be small.  In these cases, the search would be more subtle and
the constraints on VLQ can be significantly relaxed,
such that a VLQ of mass as low as 400 GeV might evade the LHC constraints.

\section{Conclusions}
The hint of a potential 750 GeV particle observed by ATLAS and CMS 
is very intriguing. At the surface value of the large production cross section, 
it is hard to interpret it in the conventional Higgs extension models, such as 
2HDMs or MSSM. However, if the additional  particles exist, e.g. vector-like 
fermions which are allowed to run in the  $H_2 \gamma\gamma$ and
$H_2 gg$ vertex, it is possible to explain the  large cross section and relatively 
large total width of the particle.

In this work, we have investigated the models with a singlet scalar that has
renormalizable couplings to the  vector-like leptons and quarks,  
taking fully account of the doublet-singlet mixing.  We have used the allowed 
parameter space regions that we obtained in recent global fits 
to the Higgs boson data. In the allowed space, we actually find 
solutions to the 750 GeV boson with $|\sin\alpha|\lsim 0.1$, 
$\Gamma_{H_2}\sim \Gamma(H_2 \to H_1H_1) + 
\Delta\Gamma^{H_2}_{\rm inv}\sim 40~{\rm GeV}$,
and $|S^{g(Q)}_{H_2}S^{\gamma(F)}_{H_2}|\sim 90$.
It remains to be seen if this excess will survive more data accumulation in 
the near future. 
Should the fitted cross section from the LHC experiments
increases or decreases in the future, we can simply modify
the product $|S^{g(Q)}_{H_2} S^{\gamma (F)}_{H_2} |$ to
fit to it.
If the 750 GeV excess turns out to be a new particle, 
new vector-like fermions may accompany and could be of utterly importance  
at the LHC Run II.

As shown in this work,
when the total decay width of the 750 GeV di-photon resonance is sizable,
it would decay dominantly into invisible particles, 
which could give rise to monojet
events with an addition gluon radiated from the initial-state gluons.
Monojet events have been searched actively at the LHC, e.g.,
at $\sqrt{s}=13$ TeV \cite{monojet-13} and at $\sqrt{s}=8$ TeV \cite{monojet-8} 
by ATLAS (CMS has similar results), in which the 95\% CL upper limits on
monojet production cross sections due to DM are given.
Let us focus on the 13 TeV data and, to be more specific, 
on a particular selection cut -- IM1
($\not\!\!\!{E}_T > 250$ GeV and $P_{T_j} > 250$ GeV). It gives an upper limit
of $\sigma \times {\rm Acceptance} \times  {\rm Efficiency} = 553$ fb.
On the other hand, the production cross section of $H_2$ via gluon fusion
is $\sigma (gg \to H_2) \sim 10^4$ fb, see the first equation in
Eq.~(\ref{eq:numeric}).
In order to radiate an additional energetic gluon from the initial-state
gluons, the cross section would decrease by a factor of $\alpha_s / 2\pi
\sim 10^{-2}$. Therefore, we expect a cross section of order $10^2$ fb for
monojet production which is obviously below the current
experimental upper limit.  
We find that the case at 8 TeV would be similar. 
Therefore, the current production
of $H_2$, which decays dominantly into DM, would still be consistent
with the monojet searches at the LHC
\footnote{
Also this channel is dependent on the UV completion of DM models as well as DM being
scalar,
fermion or vector boson for scalar mediator cases (for example, see
Ref.~\cite{Baek:2015lna}
for the Higgs portal DM case).}.

\section*{Acknowledgments}
This work is supported in part by National Research Foundation of Korea 
(NRF) Research Grant NRF-2015R1A2A1A05001869, and by the NRF grant 
funded by the Korea government (MSIP) (No. 2009-0083526) through Korea 
Neutrino Research Center at Seoul National University (PK).
%
%
The work of K.C. was supported by the MoST of Taiwan under Grants 
No. NSC 102-2112-M-007-015-MY3.


\end{document}